\documentclass[aps,prl,twocolumn,floatfix,showpacs]{revtex4-1}
\usepackage{graphicx}
\usepackage{dcolumn}
\usepackage{bm}
\usepackage{color}
\usepackage{natbib}
\usepackage{array}
\usepackage{subfigure}
\usepackage{amssymb}
\usepackage{amsbsy}
\usepackage[normalem]{ulem}
\begin{document}
    \title{Lagrangian acceleration of passive tracers in statistically-steady rotating turbulence}
    \author{Lorenzo Del Castello}
    \author{Herman J.H. Clercx}
    \affiliation{Department of Physics and J.M. Burgers Centre for Fluid Dynamics, Eindhoven University of Technology, P.O. Box 513, 5600 MB Eindhoven, The Netherlands}
    \date{\today}
    \begin{abstract}
      The statistical properties of the Lagrangian acceleration vector of passive tracers in statistically-steady rotating turbulence is studied by Particle Tracking Velocimetry. Direct effects of the background rotation are the suppression of high-acceleration events parallel to the (vertical) rotation axis, the enhancement of high-acceleration events for the horizontal acceleration, and the strong amplification of the auto-correlation of the acceleration component perpendicular to both the rotation vector $\Omega$ and local velocity vector ${\bf{u}}$. The auto-correlation of the acceleration component in the plane set up by $\Omega$ and ${\bf{u}}$ is only mildly enhanced.
    \end{abstract}
    %
    %\pacs{47.27.-i,47.32.Ef}
    %
    \maketitle
    Geophysical flows in the oceans, the atmosphere and the liquid core of the Earth, astrophysical flows, and flows inside rotating industrial machineries are all strongly affected by background rotation. Moreover, these flows are turbulent. How the Coriolis force contributes to the statistical anisotropy of rotating turbulence and its transport properties is still in debate, and the Lagrangian description of rotating turbulence has hardly been addressed. This Letter focuses on the latter aspect.\\
    Although in recent years substantial experimental~\cite{voth1998pof,mordant2001prl} and numerical~\cite{yeung1997pof} data on acceleration statistics became available for non-rotating turbulence, much less is up to date known for rotating turbulence. The present experimental study reports on the influence of the Coriolis force on the acceleration of passive tracers in statistically-steady rotating turbulence, both in terms of particle acceleration magnitude and of the Lagrangian auto-correlation of its components. The experimental data reveal the dynamical properties of the turbulent rotating flow, clearly identify the specific acceleration component affected by rotation, and provide input for future numerical and theoretical studies. The analysis is based on the same dataset as used previously to investigate Lagrangian velocity statistics~\cite{delcastello2011}, which thus focused on the Lagrangian kinematics of the flow field.
    From that study it was concluded that the velocity component parallel to the (vertical) rotation axis gets strongly reduced (compared to the horizontal ones) while rotation is increased. Moreover, the auto-correlation coefficients of the velocity components are progressively enhanced for increasing rotation rates, although the vertical one shows a tendency to decrease for slow rotation rates.\\
    The experimental setup consists of a tank filled with an electrolyte solution (density $\rho_{f}=1.19~\mathrm{g/cm^3}$ and kinematic viscosity $\nu=1.319~\mathrm{mm^2/s}$), a turbulence generator, and an optical measurement system. Four digital cameras (Photron FastcamX-1024PCI) acquire images of the central-bottom region of the flow domain. These key elements are mounted on a rotating table, so that the flow is measured in the rotating frame of reference. Technical details of the setup and turbulence generation mechanism are provided in Refs.~\cite{delcastello2011,bokhoven2009pof,delcastello2010phd}.\\ 
    The Lagrangian correlations are measured by means of Particle Tracking Velocimetry (code by ETH, Z\"{u}rich~\cite{willneff2002istpdrm}). PMMA particles (diameter $d_p=127\pm3~\mathrm{\mu m}$ and density $\rho_p=1.19~\mathrm{g/cm^3}$) are used, which are passive flow tracers both in terms of buoyancy ($\rho_p/\rho_f=1$) and inertial effects (Stokes number $St={\mathcal{O}}(10^{-3})$). The data is then processed in the Lagrangian frame, where the trajectories are filtered and the three-dimensional (3D) time-dependent signals of acceleration are extracted following the approach by L\"{u}thi~\cite{luethi2002phd}: the raw position signal at time $t$ is filtered by fitting cubic polynomials along the trajectories to remove the measurement noise. The fit is performed on a segment of the trajectory $[t-10\Delta t; t+10\Delta t]$, with $\Delta t$ the PTV time-step.
    This segment is found to be the optimal filter width to remove the background noise (with amplitude ${\mathcal{O}}(10^{-5})$ m) from the data (Kolmogorov length scale typically not smaller than 0.5 mm). For each time $t$ and each Cartesian component, a system of equations based on 21 data points is formulated and solved, yielding the coefficients of the cubic polynomials. These are then used to find the filtered position, velocity and acceleration components. The frequency response of the applied smoothing filter is proportional to $1/\Delta t^3$, $1/\Delta t^2$, and $1/\Delta t$ for $x_i(t)$, $v_i(t)$ and $a_i(t)$, respectively. With the present setup, up to 2500 particles per time-step have been tracked on average in a volume with size $100\times 100\times 100~{\rm{mm}}^3$, thus roughly $1.5{\mathcal{L}}^F$, with ${\mathcal{L}}^F=70$ mm the integral scale of the turbulent flow, along each coordinate direction.\\
    The flow is subjected to different background rotation rates $\Omega\in\lbrace0; 0.2; 0.5; 1.0; 2.0;$ $5.0\rbrace~\mathrm{rad/s}$ around the vertical $z$-axis. The PTV time-step is chosen to be $\Delta t = 16.7~\mathrm{\mu s}$ for the first four runs, and $\Delta t = 33.3~\mathrm{\mu s}$ for $\Omega=2.0~\mathrm{and}~5.0~\mathrm{rad/s}$. The mean kinetic energy of the turbulent flow is statistically steady in time and decays in space along the upward vertical direction. The flow is fully turbulent in the bottom region of the container where the measurement domain is situated. Eulerian characterisation of the (rotating) turbulent flow with stereo-PIV measurements has been reported elsewhere~\cite{bokhoven2009pof} and are in excellent agreement with data extracted from the current PTV measurements~\cite{delcastello2010phd,delcastello2011}.
    The stereo-PIV experiments reveal that the flow is statistically homogeneous in the horizontal plane and approximately statistically isotropic at midheight in the measurement domain. Statistically averaged data from the $x$- and $y$-components of the acceleration vector should yield similar results, and we will therefore consider horizontally averaged quantities only. For example, the three standard deviations for the acceleration components, $\sigma_i=\langle a^2_i\rangle ^{1/2}$, with $i\in\{x,y,z\}$, are reduced to $\sigma_h=\frac{1}{2}(\sigma_x+\sigma_y)$ and $\sigma_z$. The root-mean-square velocity $u_{rms}$ averaged over horizontal planes is typically 10 to $15~\mathrm{mm/s}$. For the Kolmogorov length and time scales we found the typical values $0.6~{\rm{mm}}\lesssim\eta\lesssim0.8~{\rm{mm}}$ and $0.25~{\rm{s}}\lesssim\tau_{\eta}\lesssim0.55~{\rm{s}}$, respectively. The Taylor-scale Reynolds number is in the range $70\lesssim Re_{\lambda}\lesssim110$ for all rotation rates. The acceleration standard deviation and the kurtosis are given in Table~\ref{tab1}.
    The strong suppression of vertical accelerations at $\Omega=5.0~{\rm{rad/s}}$, see Table~\ref{tab1}, represents a classical signature of fast rotation, {\it{i.e.}} the two-dimensionalisation of the flow field. Despite the anomalous behaviour for $\Omega=2$ rad/s (for a brief discussion, see Ref.~\cite{delcastello2011}), the ratio of vertical versus horizontal standard deviation, $\xi=\sigma_z/\sigma_h$, decreases monotonously with increasing $\Omega$.
    \begin{figure}
		\includegraphics[width=0.48\columnwidth]{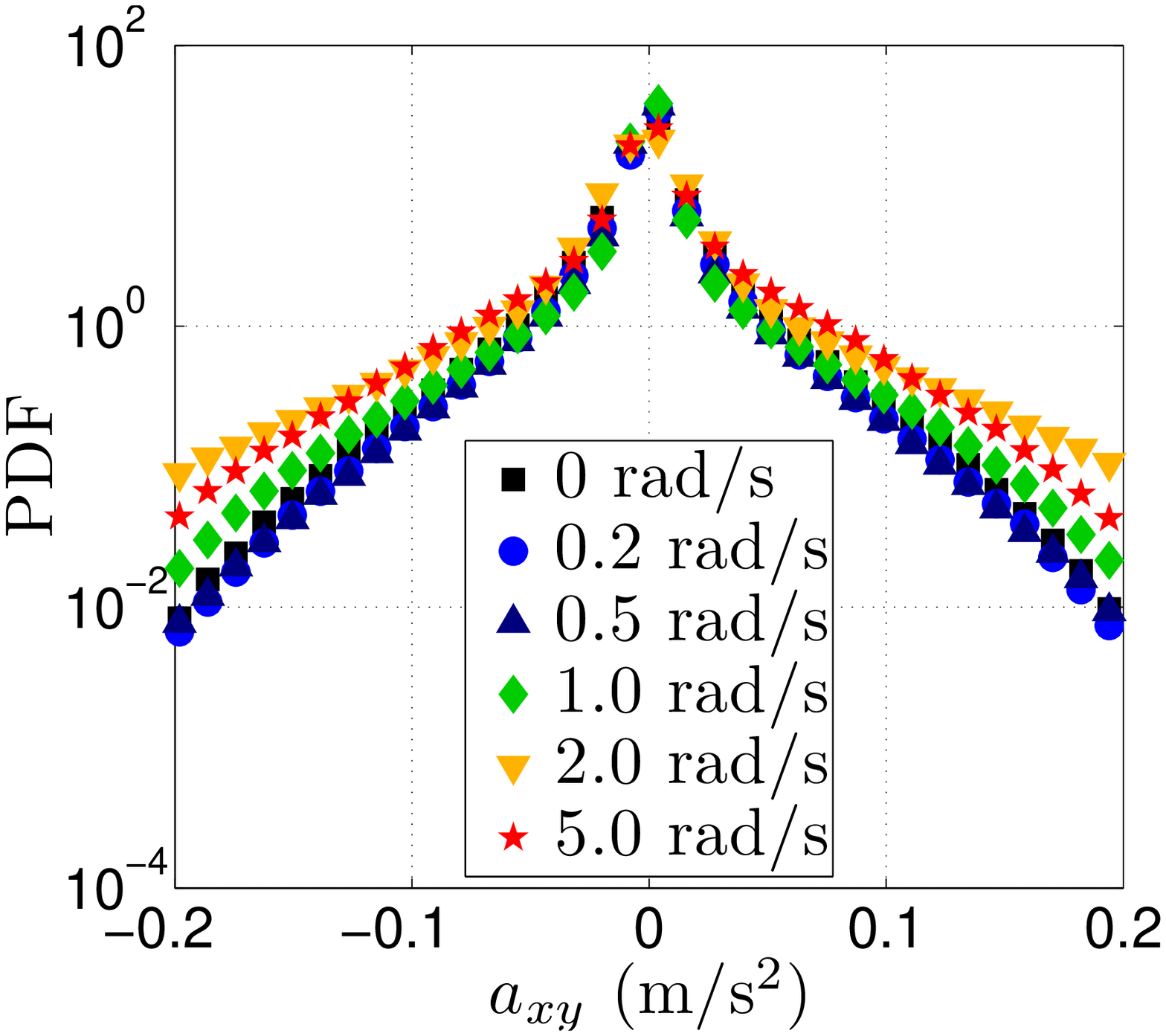}%exps123456_PDFax-fstep5-linlog.eps
		\includegraphics[width=0.48\columnwidth]{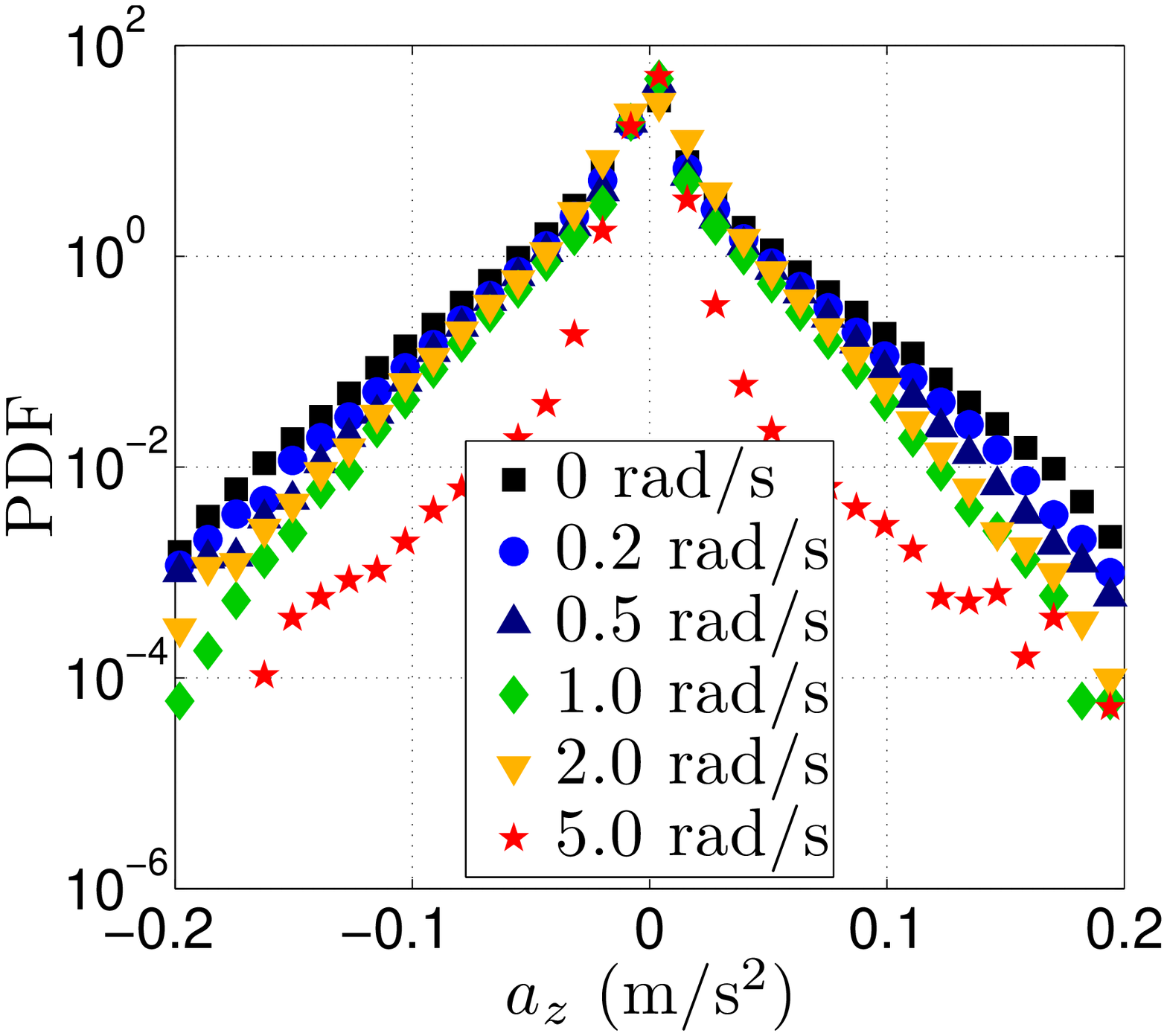}%exps123456_PDFaz-fstep5-linlog.eps
		\put(-230.0,0.0){(a)}
		\put(-112.0,0.0){(b)}
		\caption{(color online) PDFs of $a_{xy}$ (a) and $a_z$ (b) of the acceleration for all experiments in linear-logarithmic scale. The time-lag corresponds to the PTV time-step $\Delta t$ (see text).}
		\label{fig1}
    \end{figure}\\
    \noindent The influence of rotation is first analysed in terms of the PDFs of the components of the particle acceleration vector, see Fig.~\ref{fig1}. The horizontally averaged one ($a_{xy}$) is based on $8\times 10^6$ and the $z$-component on $4\times 10^6$ data points. We compared the PDFs of the acceleration components for the non-rotating experiment with results from the literature~\cite{voth1998pof}, and observed largely similar features: highly non-Gaussian distributions. In order to assess the convergence of the acceleration PDFs we performed a more systematic statistical analysis of the raw velocity data from PTV. Following Ref.~\cite{mordant2001prl}, we computed a series of PDFs of the Lagrangian velocity increments by decreasing the time-lag from $32\Delta t$ to $\Delta t$.
    We normalized the PDFs with their variances and observed the evolution from a roughly Gaussian shape to the shape of our Lagrangian acceleration PDFs for decreasing time-lag (and approaching a converged shape with decreasing time-lag). These results also support the statistical intermittency of the acceleration PDFs measured in our experiment. However, due to a slight temporal under-resolution of our measurements, we are not able to measure the highest acceleration events.
    This is revealed by the end tails of the PDFs, which gets slightly lower for accelerations higher than $0.1~\mathrm{m/s^2}$. Fortunately, this does not hamper the qualitative comparison of PDFs obtained for different rotation rates, which is the central issue in the current investigation.\\
    Rotation does not influence the PDF of the horizontal acceleration components in a monotonic way. The tails of the PDFs get slightly lower for $\Omega=0.2$ and $0.5~\mathrm{rad/s}$. They get higher and significantly higher for $\Omega=1.0$ and $2.0~\mathrm{rad/s}$, respectively. Only the end tails get slightly lower when the rotation rate is further increased from $2.0$ to $5.0~\mathrm{rad/s}$. The PDF of the vertical acceleration component, on the contrary, have its tails monotonically lowered as the rotation rate is increased. This indicates the importance of the two-dimensionalisation process induced by rotation which affects the accelerations of passive tracers, despite the same 3D steady forcing is applied to the flow at every rotation rate.
    \linespread{1.3}
    \begin{table}
		\begin{center}
			\begin{tabular}{cccccccc}
				\hline
				$\Omega$ \footnotesize{(rad/s)} & & 0 & 0.2 & 0.5 & 1.0 & 2.0 & 5.0\\
				\hline
				\hline
				$\sigma_h=\frac{1}{2}(\sigma_x + \sigma_z)$ \footnotesize{($\mathrm{mm/s^2}$)}
					&  & 28.6 & 25.3 &  24.7 & 29.4 & 41.0 & 39.3 \\
				$\sigma_z$ \footnotesize{($\mathrm{mm/s^2}$)}
					& & 24.3 &  20.6 &  18.6 &  15.9 &  19.6 &   7.1 \\
				$\xi = \sigma_z/\sigma_h $ (--)	&  & 0.85 &  0.81 &  0.75 &  0.54 &  0.48 &   0.18 \\ 
				\hline
				$K_h=\frac{1}{2}(K_x + K_y)$ \footnotesize{(--)}
					& &  11.1 &  13.8 & 15.0 & 13.6 &  8.7 &  7.9 \\
				$K_z$ \footnotesize{(--)}
					& &  10.4 &  12.6 & 13.4 & 14.9 &  9.0 & 26.7 \\
				\hline
				$Ro = u_{rms}/(2\Omega{\mathcal{L}}^F)$ \footnotesize{(--)}
					& & $\infty$ & 0.47 & 0.20 & 0.13 & 0.09 & 0.02 \\
				$Ek\times 10^5$ \footnotesize{(--)}
					& & $\infty$ & 10 & 4 & 2 & 1 & 0.4 \\
				$\delta_{Ek}$ \footnotesize{($\mathrm{mm}$)}
					& & $\infty$ & 2.5 & 1.6 & 1.1 & 0.8 & 0.5 \\
				\hline
				\hline
			\end{tabular}
			\linespread{1}
			\caption{Standard deviation $\sigma_i=\langle a_i^2\rangle ^{1/2}$ and kurtosis $K_i=\langle a_i^4\rangle/\langle a_i^2\rangle^2$ (with $i\in\{x,y,z\}$) of the acceleration distributions for all (non-)rotating experiments. For each one, the Ekman number $Ek = \nu/(\Omega L_z^2)$, with $L_z = 250~{\rm{mm}}$ the vertical size of the flow domain, and the thickness of the Ekman boundary layer $\delta_{Ek} = \sqrt{\nu/\Omega}$ is also given.}
			\label{tab1}
		\end{center}
    \end{table}\linespread{1}\\
    \noindent The PDFs shown in Fig.~\ref{fig1} are quantified by extracting values of $\sigma_i$, skewness $S_i=\langle a_i^3\rangle/\langle a_i^2\rangle^{3/2}$ and kurtosis $K_i=\langle a_i^4\rangle/\langle a_i^2\rangle^2$. As can be conjectured from Fig.~\ref{fig1}, the PDFs are not appreciably skewed, which is confirmed by the fact that $S_i\approx 0$ for all rotation rates. The values for $\sigma_i$ and $K_i$ (with $i=h$ or $z$) are presented in Table~\ref{tab1}. Although a slight decrease of $\sigma_i$ is observed for slow rotation ($\Omega\in[0.2;0.5]~\mathrm{rad/s}$), $\sigma_h$ increases substantially for large rotation rates ($\Omega\in[1.0;5.0]~\mathrm{rad/s}$).
    However, $\sigma_z$ is strongly suppressed. The values for $K_h$ and $K_z$ reveal non-monotonic variations. In fact, a mild background rotation ($\Omega\in[0.2;0.5]~\mathrm{rad/s}$) is seen to amplify the kurtosis of all acceleration components, while a further increase of rotation ($\Omega\in[1.0;2.0]~\mathrm{rad/s}$) induces a reduction of the kurtosis. Such a reduction proceeds when $\Omega$ is raised to $5.0~\mathrm{rad/s}$ for what concerns $K_h$, but $K_z$, instead, is strongly enhanced for the fastest rotating run, reflecting the strong suppression of $a_z$ induced by rotation (and quantified by the corresponding value of $\sigma_z$ in Table~\ref{tab1}).
    The values for $K_i$ for no or mild background rotation are also in good agreement with the ones reported in the literature for isotropic turbulence at comparable $Re_{\lambda}$ (see, e.g., the inset of Fig.~2(a) in Bec \textit{et al.}~\cite{bec2006jfm}).
    \begin{figure}
		\includegraphics[width=0.43\columnwidth]{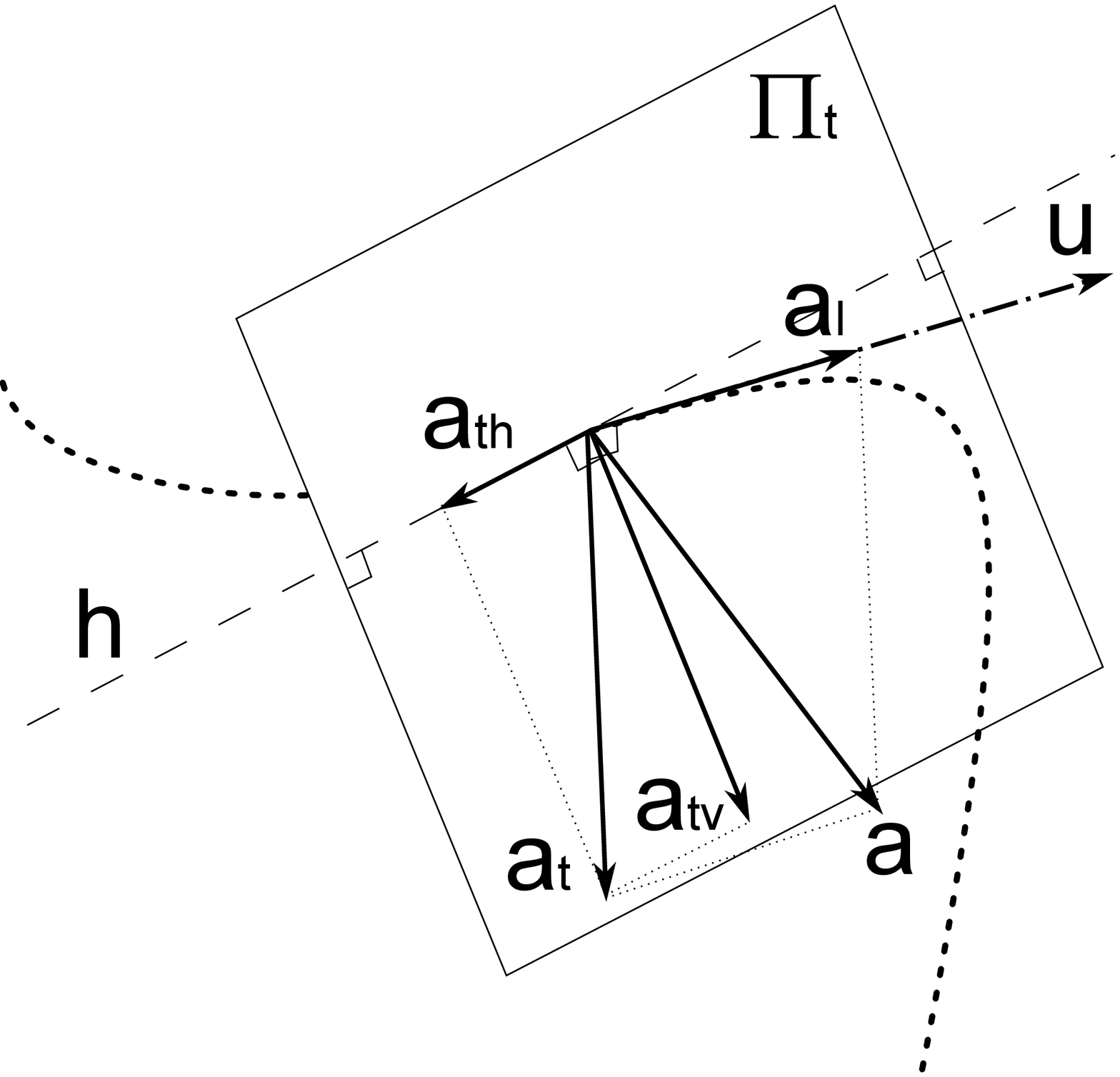}
		\includegraphics[width=0.55\columnwidth]{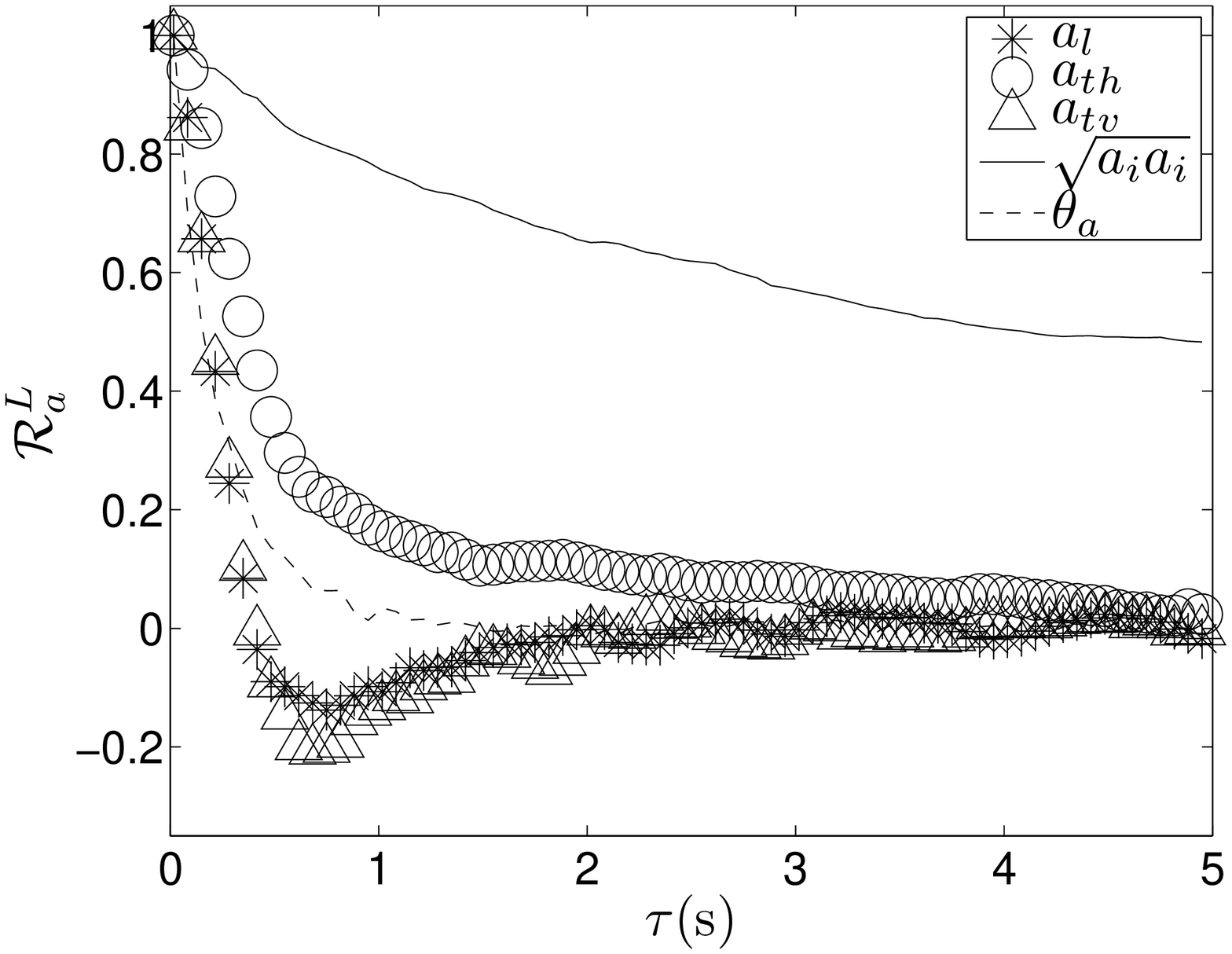}
		\put(-230.0,0.0){(a)}
		\put(-112.0,0.0){(b)}
		\caption{(a) Sketch of the decomposition of the acceleration in the longitudinal ($a_l$), transversal (partially) vertical ($a_{tv}$), and transversal horizontal ($a_{th}$) components. (b) Lagrangian auto-correlation coefficients for the non-rotating experiment. Correlations of $a_l$, $a_{tv}$, and $a_{th}$, the modulus of acceleration $|{\bf{a}}|$, and its polar angle $\theta_a$ in the horizontal $xy$-plane.}
		\label{fig2}
    \end{figure}\\
    \noindent The Lagrangian auto-correlation coefficients (as function of the time separation $\tau$) for the Cartesian acceleration components $a_i$, with $i\in\{x,y,z\}$, are obtained averaging over a sufficient number of trajectories, and normalising with the variance of the single component, \textit{i.e.} $\mathcal{R}^L_{ai}(\tau)\equiv\langle a_i(t)a_i(t+\tau)\rangle/\langle a_i^2(t)\rangle$. The Lagrangian acceleration for the non-rotating case is found to decorrelate with itself within $2.5\tau_{\eta}$, and each component shows the well-known negative loop (a mild anti-correlation at short times). The decorrelation process of the Cartesian components is due to the change of the direction of the acceleration vector, rather than to a change of its magnitude. These observations agree with known features of the Lagrangian acceleration in homogeneous isotropic turbulence.
    The values reported in the literature show a strong dependence with the Reynolds number: the time separation of the first zero-crossing of the auto-correlations of single acceleration components ranges from $2\tau_{\eta}$ to $10\tau_{\eta}$, for $100\lesssim Re_{\lambda}\lesssim 1000$. Our measurements of the non-rotating flow confirm the general picture: the dynamics of the Lagrangian acceleration vector involves both the dissipative scale $\tau_{\eta}$ (over which it rapidly changes direction), and the integral time-scale (relevant for the evolution of its magnitude).\\
    We also computed the correlation coefficients of the longitudinal ($a_l$), the transversal horizontal ($a_{th}$), and the transversal (partially) vertical ($a_{tv}$) components of the acceleration vector. This decomposition is sketched in Fig.~\ref{fig2}a, where a curved particle trajectory is marked as a thick dotted line, and the transversal plane (the plane perpendicular to the velocity vector ${\bf{u}}$) is denoted as $\Pi_t$. The acceleration vector ${\bf{a}}$ is first decomposed into its longitudinal and transversal components, where the longitudinal acceleration is defined as the projection over the velocity unit vector (${\hat{\bf{u}}}\equiv {\bf{u}}/|{\bf{u}}|$), $a_l={\bf{a}}\cdot {\hat{\bf{u}}}$.
    The transversal horizontal acceleration is defined as the projection over the direction ${\bf{h}}$, which is simultaneously perpendicular to the velocity vector ${\bf{u}}$ and to the vertical unit vector ${\bf{e}}_z$: $a_{th}={\bf{a}}\cdot {\bf{h}}$, with ${\bf{h}}\cdot {\bf{u}}=0$, ${\bf{h}}\cdot {\bf{e}}_z=0$, and $|{\bf{h}}|=1$. The direction of ${\bf{h}}$ is sketched as a thin dashed line, and it represents the intersection of the plane $\Pi_t$ with the horizontal plane passing by the current particle position. The transversal (partially) vertical acceleration is defined as the remaining component, $a_{tv}=|{\bf{a}}-a_l{\hat{\bf{u}}}-a_{th}{\bf{h}}|$, and is in general not purely vertical.
    We are particularly interested in such decomposition of the acceleration vector because the Coriolis acceleration introduced by the background rotation acts solely in the direction perpendicular to the rotation axis, and perpendicular to the velocity vector. Therefore, it is expected that rotation will directly affect the transversal horizontal component $a_{th}$ of the acceleration of the fluid particles, but it is yet unclear if and how strong the components $a_{l}$ and $a_{tv}$ are affected.
    \begin{figure}[!ht]
		\includegraphics[width=0.70\columnwidth]{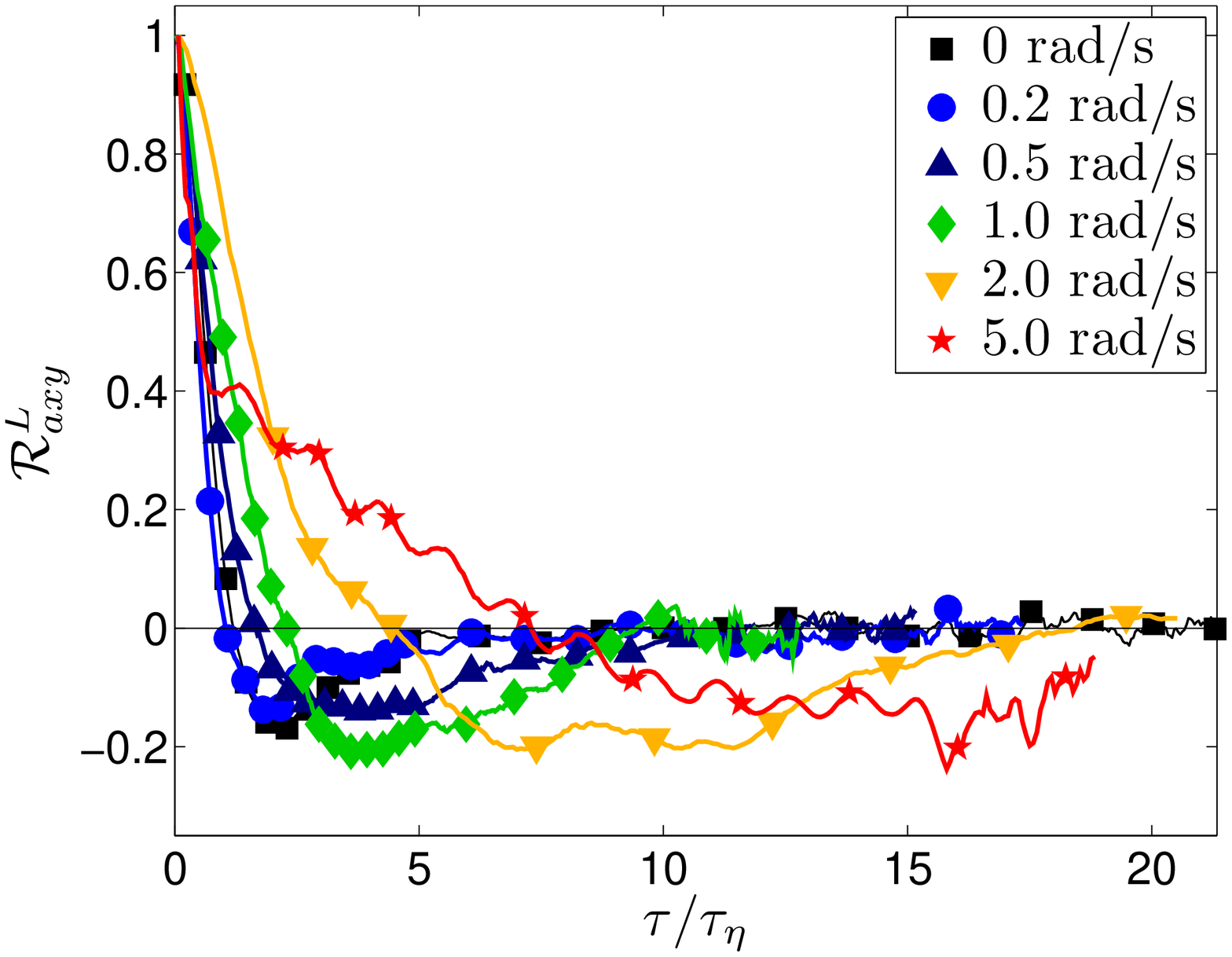}\\
		\includegraphics[width=0.70\columnwidth]{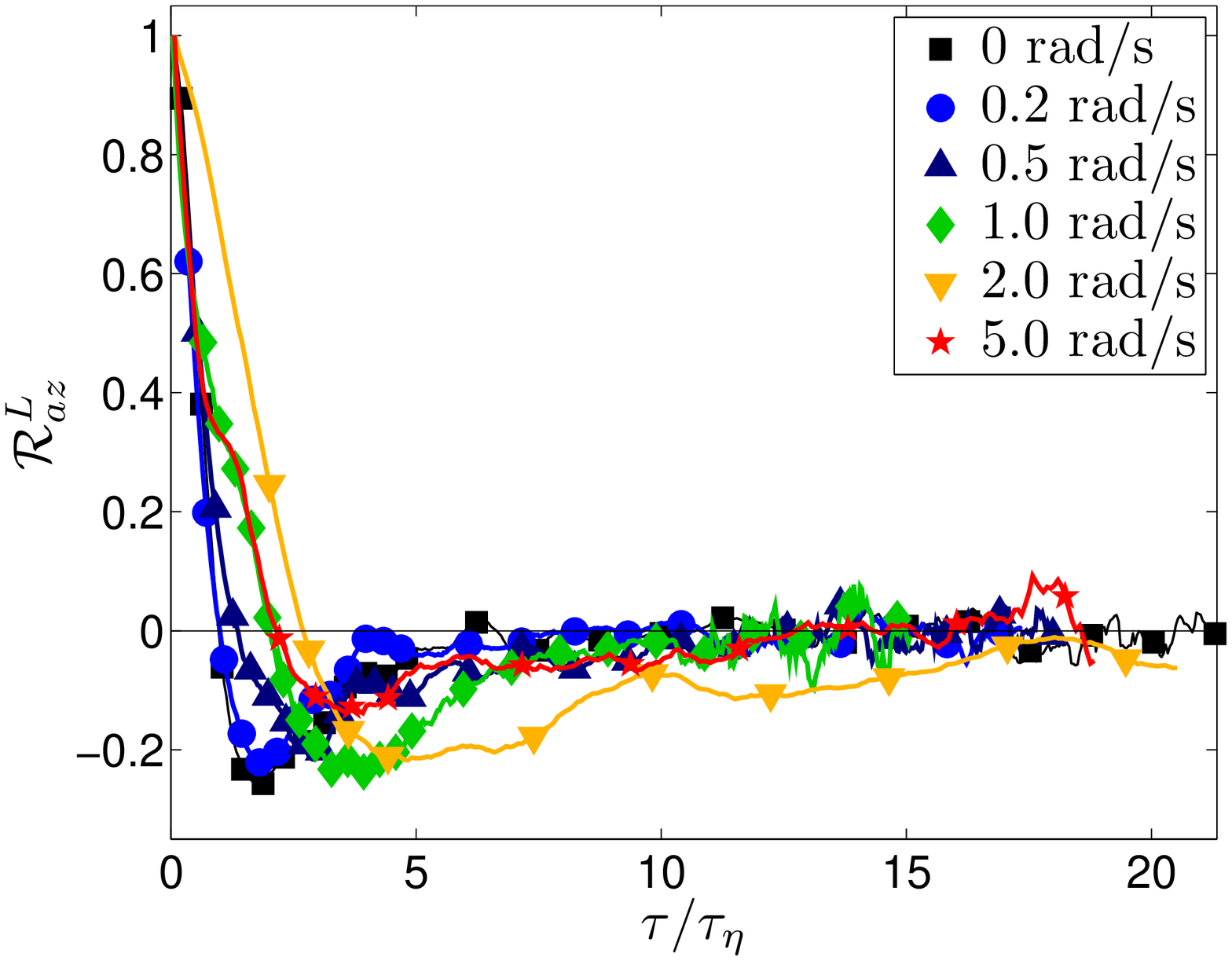}
		\caption{(Color online) Lagrangian auto-correlation coefficients of $a_{xy}$ (top) and $a_z$ (bottom) for all experiments. The time is normalised with the Kolmogorov time scale $\tau_{\eta}$.}
		\label{fig3}
	\end{figure}
    \\
    \begin{figure*}
		\includegraphics[width=0.32\textwidth]{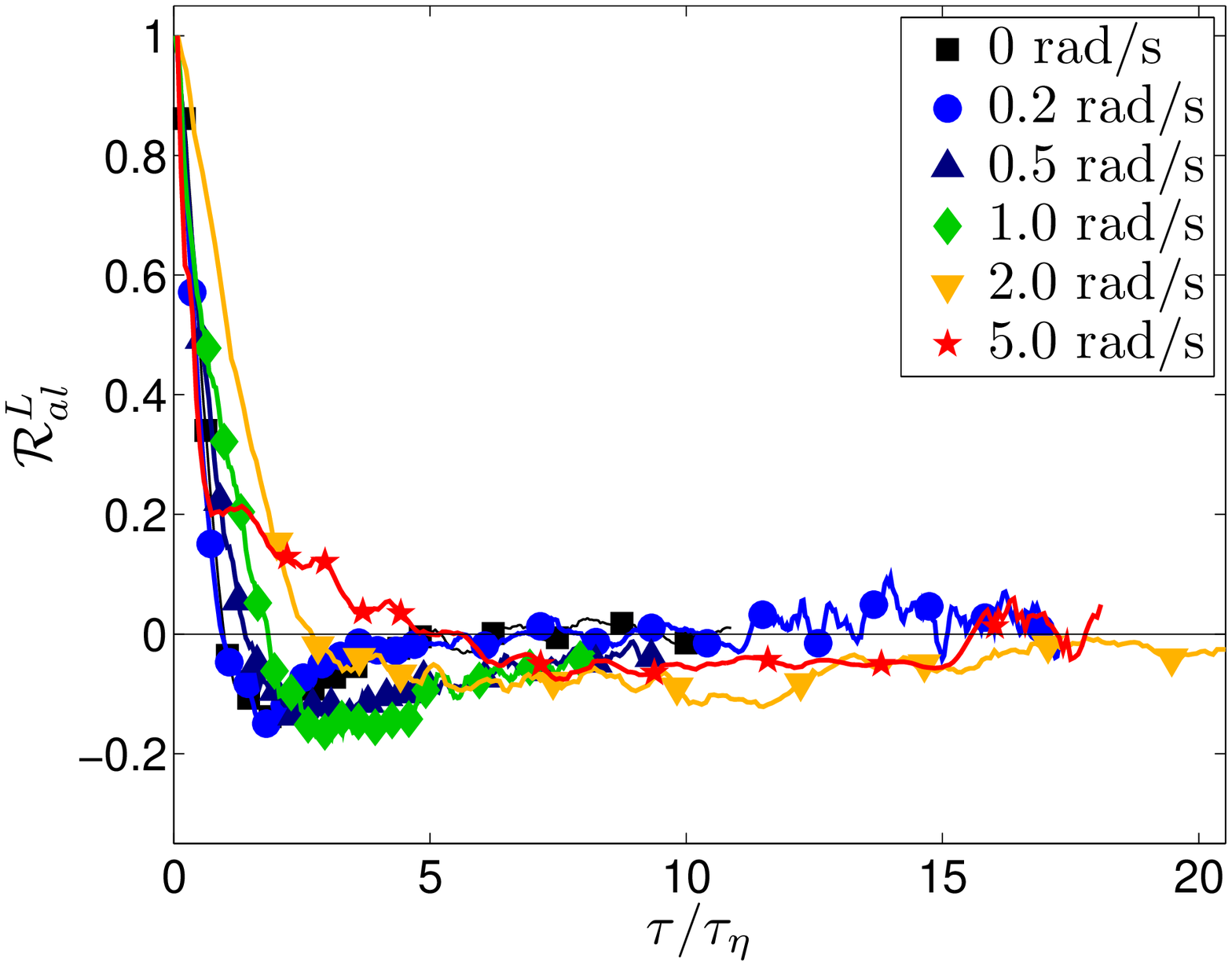}
		\includegraphics[width=0.32\textwidth]{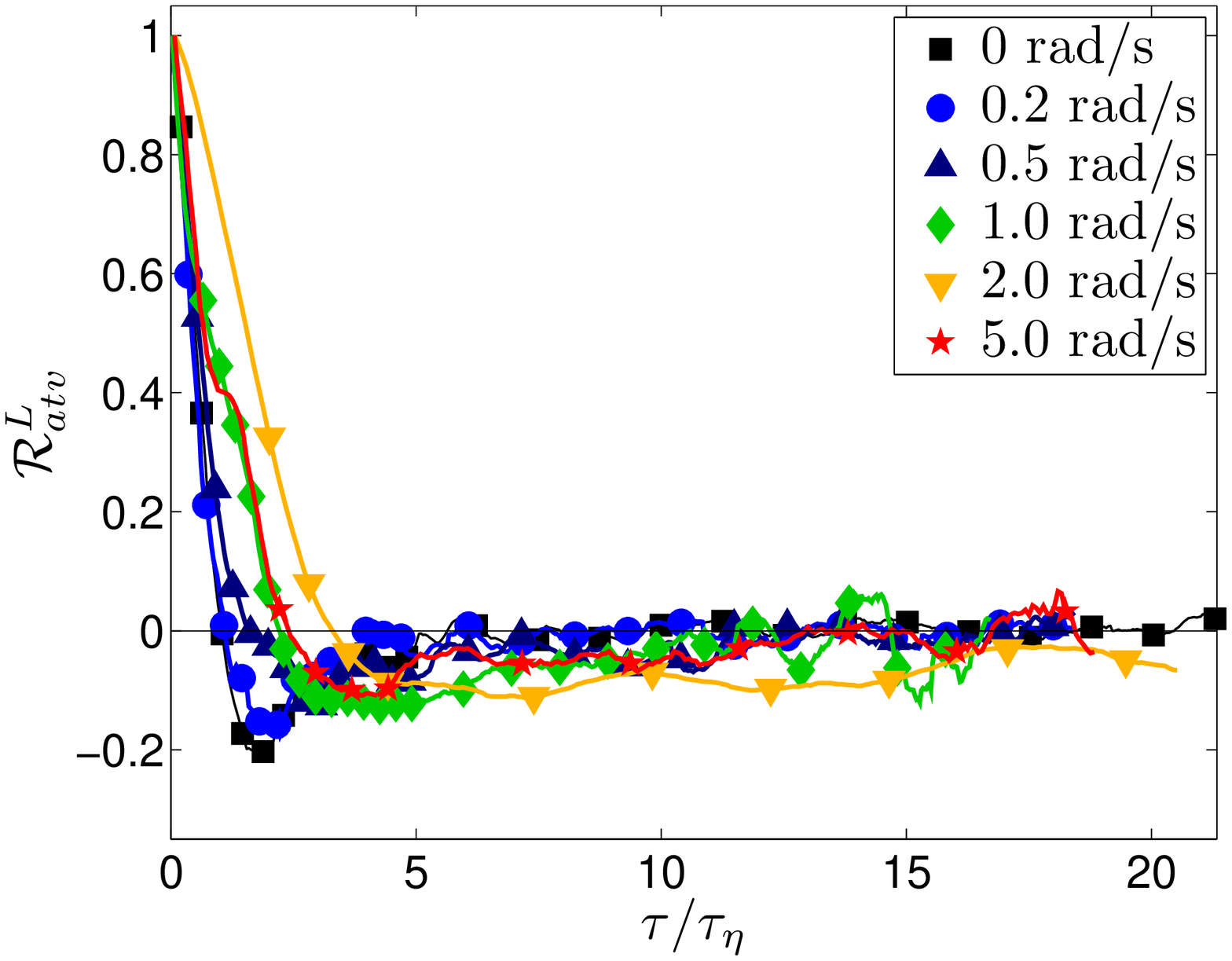}
		\includegraphics[width=0.32\textwidth]{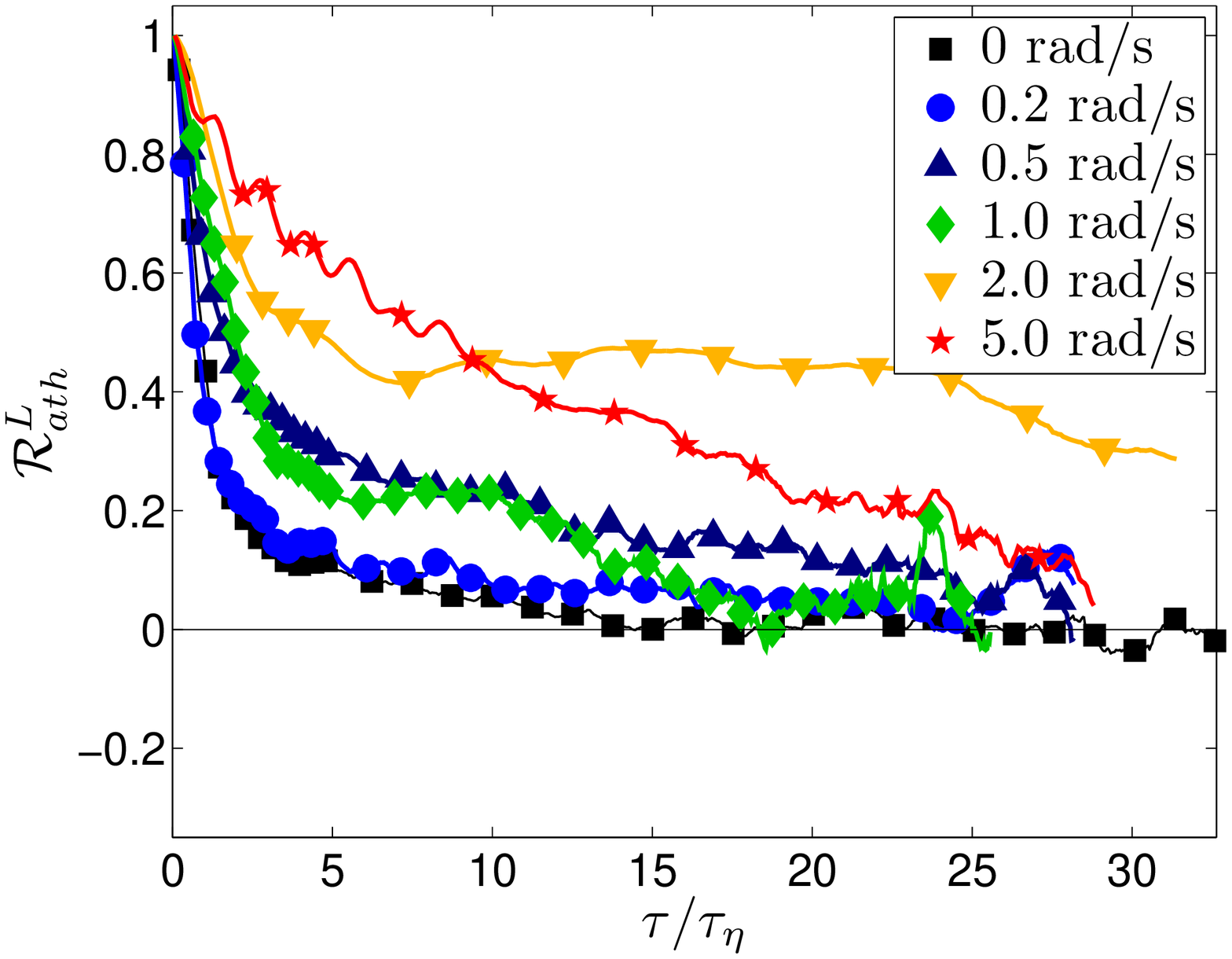}
		\caption{(Color online) From left to right, the Lagrangian auto-correlation coefficients of $a_l$, $a_{tv}$, $a_{th}$ for all experiments. The time is normalised with the Kolmogorov time scale $\tau_{\eta}$. Note that in the last plot a different scale for the time axis is used.}
		\label{fig4}
    \end{figure*}\\
    \noindent The correlation coefficients of the modulus $|{\bf{a}}|$, the polar angle $\theta_a$ in the horizontal $xy$-plane of the acceleration vector, together with the correlation coefficients of the longitudinal ($a_l$), the transversal horizontal ($a_{th}$), and the transversal (partially) vertical ($a_{tv}$) components of the acceleration vector, all for the non-rotating experiment, are shown in Fig.~\ref{fig2}b. The components $a_l$ and $a_{tv}$ decorrelate with themselves on the same (very short) time scale as that for the single Cartesian components $a_i$ and the polar angle $\theta_a$. They show a similar negative loop typical of the correlation curves of every Cartesian component. The transversal horizontal component $a_{th}$ remains mildly correlated for a longer time, roughly $5\tau_{\eta}$.\\
    The Lagrangian auto-correlation coefficients of the Cartesian acceleration components $a_h$ and $a_z$ are shown in Fig.~\ref{fig3} for all (non-)rotating experiments, with time normalised with the Kolmogorov time scale $\tau_{\eta}$ and in linear-linear scale. Fig.~\ref{fig4} displays the Lagrangian auto-correlation coefficients of the components $a_l$, $a_{tv}$, and $a_{th}$. In our data, the statistical accuracy is a decreasing function of the time-lag, and the autocorrelations reveal important statistical noise at long times. Despite this, relevant trends are clearly revealed already at short and intermediate times. The effects of rotation on the correlations of the Cartesian components get appreciable for $\Omega=1.0~\mathrm{rad/s}$, and important for $\Omega=2.0$ and $5.0~\mathrm{rad/s}$.
    For these runs, the time scale of the decorrelation process is significantly increased, as revealed by the temporal shift of the negative loop of the correlations of the horizontal component, and to a lesser extent of the vertical component. The plots shown in Fig.~\ref{fig4} illustrate that the correlations of the longitudinal and transversal (partially) vertical components are only mildly affected by rotation, even for the highest rotation rates.
    The transversal horizontal component is instead strongly affected by the background rotation: at $\Omega=2.0~\mathrm{rad/s}$ its coefficient is still around $0.3$ for time separations over $30\tau_{\eta}$, and the correlation gets only partially reduced for $\Omega=5.0~\mathrm{rad/s}$. This confirms the direct role played by the Coriolis acceleration in the amplification of the Lagrangian acceleration correlation in rotating turbulence.\\
    We have investigated the influence of the Coriolis acceleration on the statistical properties of the Lagrangian acceleration vector in statistically-steady rotating turbulence by means of PTV. Rotation is confirmed to suppress high-acceleration events (reduced intermittency) along the direction parallel to the rotation axis, and to amplify considerably the auto-correlation of the component of the transversal acceleration perpendicular to the rotation axis ($a_{th}$), while hardly affecting the other two components ($a_l$, $a_{tv}$).\\
    \textit{Acknowledgements:} This project has been funded by the Netherlands Organisation for Scientific Research (NWO) under the Innovational Research Incentives Scheme grant ESF.6239. The institutes IGP and IfU of ETH (Z\"{u}rich) are acknowledged for making available the PTV code.
\end{document}